\documentclass[prl,twocolumn,floatfix,superscriptaddress]{revtex4-1}
\usepackage{dcolumn,amsmath}
\usepackage[dvips]{graphicx}
\usepackage{bm}
\usepackage{hyperref}

\newcommand{\cm}{\ensuremath{{\rm cm}}$^{-1}$}

\newcommand{\Eref}[1]{Eq.~(\ref{#1})}

\begin{document}
\title{Calculation of the parity and time reversal violating interaction in $^{225}$RaO}

\author{A.D.\ Kudashov}\email{kudashovad.@gmail.com}
\author{A.N.\ Petrov}
\email{alexsandernp@gmail.com}
\author{L.V.\ Skripnikov}\email{leonidos239@gmail.com}
\author{N.S.\ Mosyagin}
\author{A.V.\ Titov}
\homepage{http://www.qchem.pnpi.spb.ru}
\affiliation{Institute of Physics, Saint Petersburg State University, Saint Petersburg, Petrodvoretz 198904, Russia}
\affiliation{Federal state budgetary institution ``Petersburg Nuclear Physics  Institute'', Gatchina, Leningrad district 188300, Russia}
\author{V.V.\ Flambaum}
\affiliation{School of Physics, The University of New South Wales, SYDNEY 2052
Australia}
\date{\today}

\begin{abstract}
The 10-electron generalized relativistic effective core potential and the corresponding correlation spin-orbital basis sets are generated for the Ra atom and the relativistic coupled cluster calculations for the RaO molecule are performed. The main goal of the study is to evaluate the T,P-odd parameter $X$ characterized by the molecular electronic structure and corresponding to a ``volume effect'' in the interaction of the $^{225}$Ra nucleus Schiff moment with electronic shells of RaO. Our final result for $X(^{225}$RaO$)$ is $-7532$ which is surprisingly close to that in $^{205}$TlF but has a different sign. The obtained results are discussed and the quality of the calculations is analyzed. The value is of interest for a proposed experiment on RaO [PRA 77, 024501 (2008)] due to a very large expected Schiff moment of the $^{225}$Ra nucleus.
\end{abstract}

\maketitle

\section{Introduction}

Study of fundamental interactions, which break both the time-reversal invariance~(T) and spatial parity inversion symmetry~(P), or T,P-odd interactions, is a way to study so-called ``new physics'' \cite{Ginges:04, Erler:05} beyond the Standard model of electroweak and strong interactions. Despite well known drawbacks and unresolved problems of the Standard model
(radiative corrections to the Higgs mass are quadratically divergent; rather artificial Higgs mechanism of symmetry breaking is not yet proved experimentally; the origins of CP-violation, where ``C'' is the charge conjugation symmetry, are not well understood; the CP-violation according to CPT-theorem is equivalent to T-violation problem, etc.) there are very few experimental data available which are in contradiction with this theory.

In turn, some popular extensions of the Standard model, which allow one to overcome its disadvantages, are not confirmed experimentally.
Since considerable enhancement of the T,P-violation effects is expected in polar diatomic molecules comprising a heavy atom, such systems have been a research subject for a couple of decades.
It is also believed that the Standard model gives the baryon-to-photon ratio 10 orders of magnitude smaller
than the observed value. This is a strong argument in favour of additional source of CP-violation which may be
detected using atomic and molecular experiments searching for T,P-violating interactions.
Thus, finding a molecule that exhibits the strongest T,P-odd effects may be crucial for observation of these effects experimentally.

Following \cite{Hinds:80a}, the effective interaction with the Ra nucleus Schiff moment in RaO can be written in the form
\begin{equation}
     H_{\rm eff}= 6SX \vec{\sigma}_N \cdot \vec{\lambda}\ ,
 \label{interaction}
\end{equation}
where $\vec{\sigma}_N$ is the Ra nuclear spin operator, $\vec{\lambda}$ is the unit vector along the internuclear axis $z$ from Ra to O, $S$ is the Shiff moment of Ra, $X$ is determined by the electronic structure of the molecule:
\begin{equation}
    X=\frac{2\pi}{3}\left[\frac{\partial}{\partial z}\rho_{\psi}\left(\vec{r}\right)\right]_{x,y,z=0} \label{X},
\end{equation}
where  $\rho_{\psi}\left(\vec{r}\right)$  is an electronic density calculated from the four-component wave function $\psi$. The amplitude of the T,P-odd spin-axis interaction, $6SX$, in RaO was estimated by Flambaum \cite{Flambaum:01} to be 500 times larger than that in TlF \cite{Petrov:02}. Such an enhancement in RaO is based firstly on the $^{225}$Ra nucleus Schiff moment \cite{RaSchiffmoment:01, RaSchiffmoment:02} being about 200 times greater than that of $^{205}$Tl \cite{TlSchiffmoment:01, TlSchiffmoment:02, TlSchiffmoment:03} and secondly on a semiempirical estimate $X($$^{225}$RaO$)=2.2X($$^{205}$TlF$)$.

In the paper we have performed ab initio calculations of the T,P-odd parameter $X(^{225}$RaO$)$, excitation energies of the Ra atom and some spectroscopic properties of the $^{225}$RaO molecule.

\section{Methods}

It follows from eq.~\ref{X} that the volume effect is localized on the nucleus and determined by polarization of the valence electrons. Earlier it was demonstrated by our group \cite{Titov:06amin} that calculation of such core properties can be performed efficiently in two steps. Firstly, electron correlation for valence and outer-core electrons is accounted for, whereas the inner-core electrons are excluded from this calculation using the generalized relativistic effective core potential (GRECP) method \cite{Mosyagin:10a,Titov:99}, which yields a very accurate valence region wave function by the most economical way. Secondly, since the heavy atom's inner core wave function cannot be accurately obtained within the GRECP method, it has to be recovered using a non-variational restoration procedure \cite{Titov:06amin}. The two-step approach has recently been used in \cite{Baklanov:10,Petrov:07a,Skripnikov:09,Skripnikov:11a,Petrov:11} for calculation of other core properties, such as hyperfine constants, electron electric dipole moment enhancement factor, etc., in molecules and atoms.

The fully-relativistic Fock-space coupled cluster code with single and double cluster amplitudes (FS-RCCSD) \cite{MolRCCSD, Kaldor:04ba} was applied to take account of both the electron correlation and relativistic effects. Triple cluster amplitudes and basis set enlargement corrections to $X$ were obtained using scalar-relativistic approach within {\sc cfour} \cite{CFOUR} and {\sc mrcc} \cite{Kallay:1,Kallay:3} codes via interface to our new non-variational one-center restoration  codes developed in \cite{Skripnikov:11a}.

\subsection{GRECP generation and atomic calculations}

The GRECP for Ra is generated to simulate interactions of ten explicitly treated electrons, the valence and outer-core ($6s6p7s$) ones, with the inner-core $1s-5d$ electrons, which are excluded from calculations.

Basis sets for Ra were constructed using the generalized correlated scheme \cite{Mosyagin:00}; as a result, five correlation spin-orbital basis sets were generated, which are designated here as A, B, C, D and E. Each one is constructed of optimized primitive $20s$, $20p$, $10d$, $8f$ and $5g$ Gaussian-type functions contracted to $[4s4p3d1f1g]$, $[5s6p4d3f1g]$, $[6s7p5d4f2g]$, $[6s8p4d2f1g]$ and $[6s8p5d5f3g]$ sets, respectively. These basis sets were obtained in a series of atomic two-component GRECP calculations, with correlations included by the FS-RCCSD method \cite{Kaldor:04ba}, and optimized to reproduce the $7s^2 \rightarrow 7s7p$ and $7s^2 \rightarrow 7s6d$ transition energies of the atom as accurately as possible.
    
Our FS-RCCSD results are compared to experimental data \cite{Rasmuss:01,Russell:01} in the table \ref{Ra levels}. A reasonable level of accuracy has been achieved with C and E basis sets. The use of D basis set leads to satisfactory results as far as $7s^2 \rightarrow 7s7p$ transition energies are concerned. However, a lack of $f$ and $g$ functions apparently causes considerable errors in $7s^2 \rightarrow 7s6d$ transition energies.

\begin{table}[!h]
\caption{Transition energies (TE) from GRECP/FS-RCCSD calculations of the lowest-lying states of the Ra atom for 10 correlated  electrons. All values are in cm$^{-1}$.}
\label{Ra levels}
    \begin{tabular}{llrrrrrr}
     \hline
     \hline
         Leading				  &       & \multicolumn{5}{c}{Basis set} &   Exper.\  \\
\cline{3-6}
         conf.                    & Term  & A        & B        & C        & D        & E     & \cite{Rasmuss:01,Russell:01} \\
     \hline
$7s_{1/2}^2 \rightarrow$ \\
     $7s_{1/2}^1 7p_{1/2}^1$      & (J=0) & 12435    & 12920    & 13072    & 13164    & 13029 & 13078 \\
     $7s_{1/2}^1 7p_{1/2}^1$      & (J=1) & 13386    & 13840    & 13995    & 14136    & 13953 & 13999 \\
     $7s_{1/2}^1 7p_{3/2}^1$      & (J=1) & 21546    & 21258    & 21045    & 21714    & 21000 & 20715 \\
     $7s_{1/2}^1 7p_{3/2}^1$      & (J=2) & 16262    & 16660    & 16756    & 16996    & 16714 & 16689 \\
     $7s_{1/2}^1 6d_{3/2}^1$      & (J=1) & 15579    & 14112    & 13892    & 14964    & 13753 & 13716 \\
     $7s_{1/2}^1 6d_{3/2}^1$      & (J=2) & 15969    & 14434    & 14199    & 15318    & 14049 & 13994 \\
     $7s_{1/2}^1 6d_{5/2}^1$      & (J=2) & 19864    & 17943    & 17553    & 18946    & 17340 & 17081 \\
     $7s_{1/2}^1 6d_{5/2}^1$      & (J=3) & 16901    & 15191    & 14925    & 16207    & 14761 & 14707 \\

     \hline\hline
    \end{tabular}
    \end{table}

\subsection{Molecular calculations}

To perform relativistic Fock-Space coupled cluster calculation of RaO, the basis set D was used instead of E, thus, excluding several $d$, $f$ and $g$ basis functions to reduce the computational efforts. Note that the major contribution to the $X$ value is determined by the $s{-}p$ orbital mixing, therefore, such simplification is justified here. The ($10s5p2d1f$)/[$4s3p2d1f$] basis set from the {\sc molcas} 4.1 library was used for oxygen.


A one-component self-consistent-field (SCF) calculation of the $(1\sigma \dots 5\sigma)^{10}\-(1\pi 2\pi)^{8}$ ground state of RaO is performed first.  It is followed by two-component RCC calculations taking account of single and double (RCCSD) cluster amplitudes. To estimate the convergence of the results with respect to the basis set enlargement we have performed two scalar-relativistic calculations using the coupled cluster method with single and double cluster amplitudes in the basis set D and in enlarged basis set (uncontracted basis set D) using \textsc{cfour} code \cite{CFOUR}.


Correction for the contribution of the triple cluster amplitudes was also estimated. For this we have performed (i) scalar-relativistic calculation using the coupled clusters with single and double cluster amplitudes and (ii) scalar-relativistic calculation using the coupled clusters with single, double and triple cluster amplitudes (CCSDT). The final $X$ value is obtained as
\begin{equation}
\begin{array}{l}
X({\rm FINAL}) = X({\rm RCCSD}) + X({\rm CCSD~~ uncon.})~ +\\[0.5em]
+ ~X({\rm CCSDT})-2X({\rm CCSD}).\\
\end{array}
\end{equation}
%

The GRECP/RCCSD method was also used to calculate the ground state potential curve of the RaO molecule, which was then used to obtain spectroscopic properties of RaO. The results are listed in table \ref{Props}. All molecular calculations of $X$ were carried out for the equilibrium internuclear distance, $R_e=3.852$, given also in table \ref{Props}.
    


\begin{table}[!h]
 \caption{Parameter $X$ calculated for the $^{225}$RaO ground state $R_e=3.852$ a.u. 
 Only the GRECP/RCCSD result takes into account the spin-orbit interaction. All values are in a.u.} \label{Calculated X}
    \begin{tabular}{lr|r}
     \hline
     \hline 
%
      \multicolumn{2}{l}{Method}           & {$X$}         \\
     \hline 
     \multicolumn{2}{l}{Hartree-Fock 
     }                       &  -9609         \\       
%
%
     \multicolumn{2}{l}{RCCSD}                              &  -7696         \\
     \multicolumn{2}{l}{CCSD}                               &  -7648         \\
     \multicolumn{2}{l}{CCSDT}                              &  -7209         \\
     \multicolumn{2}{l}{CCSD uncontracted}                  &  -7923         \\
     \multicolumn{2}{l}{FINAL}                              &  -7532         \\
     \hline\hline
    \end{tabular}
    \end{table}

\begin{table}[!h]
\caption{Calculated properties of RaO: equilibrium internuclear distance ($R_e$), harmonic frequency ($w_e$) and vibrational anharmonicity ($w_e x_e$).} \label{Props}
    \begin{tabular}{lr}
     \hline \hline
     $R_e$, \AA      &  3.852\\
     $w_e$, \cm      &  598\\
     $w_e x_e$, \cm  & 1.09\\
     \hline \hline
    \end{tabular}
\end{table}

\section{Results and discussions}

Although triple and higher cluster amplitudes in the valence region are important for chemical and spectroscopic properties in general, it can be concluded from table \ref{Calculated X} that GRECP/RCCSD calculation includes all major correlation contributions to $X$, since triple cluster amplitudes contribute only about $6\%$. Even smaller correction was found for basis set enlargement. We do not expect that further enlargement of the basis set and accounting for quadrupole amplitudes will change the result by more then 1\%. The influence of the inner core $-$ valence electron correlation on $X$ value was estimated in \cite{Dzuba:02} to be no more than 2\% for TlF. Taking into account essentially different electronic structures of TlF and RaO we believe that our final result for $X$ is reliably valid within 10\% of accuracy.


One of the goals of this work is to determine whether or not the RaO molecule is a feasible candidate for experimental research of T,P-odd effects. The corrected value $X($$^{225}$RaO$)\approx -X($$^{205}$TlF$)$, according to our calculations, is not as large as could be expected from the usual $Z$-scaling (see \cite{Z-scaling:01} and references). The estimation based on $Z$-scaling determines only a ``systematic'' variation with increasing $Z$, and does not take into account some particular changes from one element to the next in the Periodic table and features of chemical bonding with different elements (F in the case of TlF and O in the case of RaO). A very large estimated Schiff moment of $^{225}$Ra \cite{RaSchiffmoment:01, RaSchiffmoment:02} will strengthen the T,P-odd interactions in RaO by two orders of magnitude. The best upper bound on the nuclear T,P-odd interactions can be obtained using the limit on the $\rm ^{199}Hg$ EDM, $\rm |d(^{199}Hg)| < 3.1 \times 10^{-29} e \cdot cm $\cite{Griffith:09}, and the atomic calculation that links the atomic EDM with the Schiff moment of $^{199}$Hg nucleus \cite{Dzuba:02}, $\rm d(^{199}Hg)= -2.8 \times 10^{-17 } (S(^{199}Hg /e \, fm^3) e \cdot cm$. Combining these values, the current upper limit on $\rm |S(^{199}Hg)|$  is  $\rm 1.1 \times 10^{-12} /e \, fm^3$. 
Using the corresponding experimental datum for TlF \cite{Cho:91} and the most accurate calculated value for $X$ \cite{Petrov:02} one can obtain the upper limit on $\rm |S(^{205}Tl)|$ as $\rm 1.7 \times 10^{-10} /e \, fm^3$. Supposing the same accuracy for a ``speculative'' RaO experiment as for TlF (performed twenty two years ago), the same limit on $\rm |S(^{225}Ra)|$ will be obtained. However, taking into account the relation $\rm S(^{225}Ra)/S(^{199}Hg) \approx 200$ \cite{RaSchiffmoment:01, RaSchiffmoment:02, TlSchiffmoment:01, TlSchiffmoment:02, TlSchiffmoment:03} and the upper bound on $\rm |S(^{199}Hg)|$ we obtain $\rm |S(^{225}Ra)| < 2.2 \times 10^{-10} /e \, fm^3$ which is only slightly (1.3 times) worse than that expected from the speculative RaO experiment. However, considering the impressive progress in molecular spectroscopy during last decades one can expect an order of magnitude or even better sensitivity of the RaO experiment than that attained on TlF.

\subsection{Analysis of contributions to $X$} 

In the present paper an analysis of contributions to $X$ based on the one-center representation of the one-particle density matrix has been performed assuming the scalar-relativistic approximation used in the paper to construct initial (reference) Hartree-Fock wavefunction and molecular orbitals. The analysis is similar to that performed by our group for  Eu$^{2+}$ in Ref.~\cite{Skripnikov:11a}.
%
It should be noted that such one-center density matrix analysis is free from those rotations in the space of molecular orbitals which do not change the wavefunction of a considered state (e.g., rotations between the doubly occupied orbitals) and can also be easily performed in the case of a correlated wavefunction, i.e.\ beyond the one-configuration approximation.
A mean value of the one-electron operator $\bm{X}$ can be evaluated as follows:
\begin{equation} 
   \langle {\bm{X}} \rangle\ =\ \sum_{pq} D_{pq} 
    {X}_{qp}\ ,
 \label{<X>} 
\end{equation} 
where ${X}_{pq}$ are matrix elements of $\bm{X}$, $D_{pq}$ are matrix elements of the one-particle density matrix, and indices $p, q$ enumerate all the atomic basis functions centred on Ra and O assuming the conventional MO~LCAO approximation used in the RaO calculation. For a qualitative analysis of contributions to $X$ it is reasonable to rewrite \Eref{<X>} in terms of only \textit{atomic orbitals} of Ra, both occupied and virtual, which can be calculated using Hartree-Fock method for a single Ra atom since such an analysis can be performed excluding one more ambiguity caused by overlapping of the atomic orbitals on different centers.

Represent the molecular density matrix $D_{pq}$ in a ``sufficiently complete'' basis set of these atomic orbitals with indices $i, j$, running over all the occupied and virtual orbitals of Ra , i.e., decompose all original basis functions $p, q$ (centered either on Ra or O) into one-center functions $i, j$. Then the mean value for $X$ can be rewritten in the form
\begin{equation}
   \langle \widetilde{\bm{X}} \rangle\ =\ \sum_{ij} \widetilde{D_{ij}} 
   X_{ji}\ ,
 \label{<X_new>} 
\end{equation}
where ${X}_{ij}$ are matrix elements of $\bm{X}$ in the enlarged basis set of Ra orbitals, described above. $\langle {\bm{\widetilde{X}}} \rangle$ can be slightly different from $\langle {\bm{X}} \rangle$ due to incompleteness of the used enlarged Ra basis set. In practice however, this difference can be made negligible.

Applying this scheme to RaO we obtain the following contributions: the main one (more than 50\%) is due to a polarization of outer-core $6p_z$ orbital of Ra ($z$-axis coincides with the axis of the RaO molecule) into $7s, 8s...$. The rest are due to mixing of $7s, 8s, 9s \dots$ orbitals of Ra with different $p_z$ orbitals of Ra which partially can consist of orbitals of oxygen (in terms of the basis set $p, q$ used in RaO calculations). 

It should be noted that this kind of analysis can not be performed directly for individual contributions from \textit{canonical} molecular orbitals (see table \ref{Xmos}) due to the above mentioned large unitary rotations between atomic orbitals of radium (leading to mutually compensated unphysical contributions) and their overlapping with orbitals of oxygen. These circumstances result in the presence of several leading contributions. This is opposite to the case of TlF (see  Ref.~\cite{Petrov:02}) where, due to the relatively simple electronic structure, such orbital analysis was successfully applied with only one leading contribution.

\begin{table}[!h]
 \caption{Parameter $X$ calculated for the $^{225}$RaO ground state $R_e=3.852$ a.u. Individual shell contributions are calculated from the spin-averaged GRECP/SCF orbitals. All values are in a.u.} \label{Xmos}
    \begin{tabular}{l|r}
     \hline
     \hline
 
      Shell : main contribution           & {$X$}         \\
     \hline    
     {$1\sigma^2$ : $1s^2(O)$}            &    -50         \\
     {$2\sigma^2$ : $6s^2(Ra)$}           &   3862         \\
     {$3\sigma^2$ : $2s^2(O)$}            &   9936         \\
     {$4\sigma^2$ : $6p_z^2(Ra)$}         & -14426         \\
     {$5\sigma^2$ : $2p_z^2(O)$}          &  -9036         \\
     {$1\pi^4$ : $6p_x^2 6p_y^2(Ra)$}     &     50         \\
     {$2\pi^4$ : $2p_x^2 2p_y^2(O)$}      &     54         \\
     {Total SCF(spin-averaged)}           &  -9609         \\
     \hline\hline
    \end{tabular}
    \end{table}

\subsection{Analysis of $X$ sign}


The different sign of the $X$ in TlF and RaO molecules can be explained in a simple qualitative analysis. The valence electronic configurations of Ra and Tl are $6s^26p^67s^2$ and $6s^26p^1$, respectively. TlF has ionic bonding, the $6p$ valence orbital of thallium becomes mainly unoccupied due to interaction with fluorine. Thus, qualitatively, the electronic configurations of the TlF molecule can be written as Tl$^+$($6s^2$)F$^-$($2s^22p^6$). Electronic structure of RaO is more complicated. Calculation shows that its configuration is rather close to Ra$^+$($6s^26p^6$)($\pi^4$)O$^-$($2s^2 2p_z^2$), where $\pi$ is three quarters $ 2p_{x(y)}(\rm{O})$ and one quarter $ 6d_{xz(yz)}(\rm{Ra})$. According to the Mulliken population analysis, an effective configuration of Ra in RaO is $6s^2 6p^6 6d^1$. Thus, due to the interaction with oxygen, the $7s$ valence orbital of Ra becomes mainly unoccupied in RaO. Following the density matrix analysis the leading contribution to $X$ in TlF and RaO is due to the polarization of $6s$ and $6p$ orbitals, correspondingly (see above).

Taking account of the leading contributions to $X({\rm TlF})$ and $X({\rm RaO})$ discussed earlier we can write
\begin{eqnarray}
X({\rm TlF}) = 2<6s|V|6p><6s|W|6p>/\Delta E_{\rm TlF}, \\
X({\rm RaO}) = 2<6p|V|7s><6p|W|7s>/\Delta E_{\rm RaO},
\end{eqnarray}
where $\Delta E_{\rm TlF} = E_{6s}-E_{6p}$, $\Delta E_{\rm RaO} = E_{6p}-E_{7s}$, $V$ is the polarization operator, $W$ is an operator for the $X$ property localized on the Ra and Tl nuclei (see \Eref{X}). Note that $\Delta E_{\rm TlF}$ and $\Delta E_{\rm RaO}$ have the same signs. Let us choose the phase of orbitals so that the first extremum would have a positive sign for all the functions of Ra and Tl. Then the matrix elements $<6s|W|6p>$ and $<6p|W|7s>$, localized on the nucleus, will have the same signs. However, the~ terms $<6s|V|6p>$ and $<6p|V|7s>$, localized in the valence region, will have different signs, which explains the opposite signs of $X$(TlF) and $X$(RaO).

\section{ACKNOWLEDGMENTS}

Funding for the work at PNPI was provided by Russian Ministry of Education and Science, contract \#\,07.514.11.4141.
This work is supported by the RFBR grant 13-02-01406. L.S.\ is grateful to the Dmitry Zimin ``Dynasty'' Foundation. The molecular calculations were performed at the Supercomputer ``Lomonosov''.


\begin{thebibliography}{30}
\expandafter\ifx\csname natexlab\endcsname\relax\def\natexlab#1{#1}\fi
\expandafter\ifx\csname bibnamefont\endcsname\relax
  \def\bibnamefont#1{#1}\fi
\expandafter\ifx\csname bibfnamefont\endcsname\relax
  \def\bibfnamefont#1{#1}\fi
\expandafter\ifx\csname citenamefont\endcsname\relax
  \def\citenamefont#1{#1}\fi
\expandafter\ifx\csname url\endcsname\relax
  \def\url#1{\texttt{#1}}\fi
\expandafter\ifx\csname urlprefix\endcsname\relax\def\urlprefix{URL }\fi
\providecommand{\bibinfo}[2]{#2}
\providecommand{\eprint}[2][]{\url{#2}}

\bibitem[{\citenamefont{Ginges and Flambaum}(2004)}]{Ginges:04}
\bibinfo{author}{\bibfnamefont{J.~S.~M.} \bibnamefont{Ginges}}
  \bibnamefont{and} \bibinfo{author}{\bibfnamefont{V.~V.}
  \bibnamefont{Flambaum}}, \bibinfo{journal}{Phys.\ Rep.}
  \textbf{\bibinfo{volume}{397}}, \bibinfo{pages}{63} (\bibinfo{year}{2004}).

\bibitem[{\citenamefont{Erler and {Ramsey-Musolf}}(2005)}]{Erler:05}
\bibinfo{author}{\bibfnamefont{J.}~\bibnamefont{Erler}} \bibnamefont{and}
  \bibinfo{author}{\bibfnamefont{M.~J.} \bibnamefont{{Ramsey-Musolf}}},
  \bibinfo{journal}{Prog.\ Part.\ Nucl.\ Phys} \textbf{\bibinfo{volume}{54}},
  \bibinfo{pages}{351} (\bibinfo{year}{2005}).

\bibitem[{\citenamefont{Hinds and Sandars}(1980)}]{Hinds:80a}
\bibinfo{author}{\bibfnamefont{E.~A.} \bibnamefont{Hinds}} \bibnamefont{and}
  \bibinfo{author}{\bibfnamefont{P.~G.~H.} \bibnamefont{Sandars}},
  \bibinfo{journal}{Phys.\ Rev.\ A} \textbf{\bibinfo{volume}{21}},
  \bibinfo{pages}{471} (\bibinfo{year}{1980}).

\bibitem[{\citenamefont{Flambaum}(2008)}]{Flambaum:01}
\bibinfo{author}{\bibfnamefont{V.~V.} \bibnamefont{Flambaum}},
  \bibinfo{journal}{Phys. Rev. A} \textbf{\bibinfo{volume}{77}},
  \bibinfo{pages}{024501} (\bibinfo{year}{2008}).

\bibitem[{\citenamefont{Petrov et~al.}(2002)\citenamefont{Petrov, Mosyagin,
  Isaev, Titov, Ezhov, Eliav, and Kaldor}}]{Petrov:02}
\bibinfo{author}{\bibfnamefont{A.~N.} \bibnamefont{Petrov}},
  \bibinfo{author}{\bibfnamefont{N.~S.} \bibnamefont{Mosyagin}},
  \bibinfo{author}{\bibfnamefont{T.~A.} \bibnamefont{Isaev}},
  \bibinfo{author}{\bibfnamefont{A.~V.} \bibnamefont{Titov}},
  \bibinfo{author}{\bibfnamefont{V.~F.} \bibnamefont{Ezhov}},
  \bibinfo{author}{\bibfnamefont{E.}~\bibnamefont{Eliav}}, \bibnamefont{and}
  \bibinfo{author}{\bibfnamefont{U.}~\bibnamefont{Kaldor}},
  \bibinfo{journal}{Phys.\ Rev.\ Lett.} \textbf{\bibinfo{volume}{88}},
  \bibinfo{pages}{073001} (\bibinfo{year}{2002}).

\bibitem[{\citenamefont{Auerbach et~al.}(1996)\citenamefont{Auerbach, Flambaum,
  and Spevak}}]{RaSchiffmoment:01}
\bibinfo{author}{\bibfnamefont{N.}~\bibnamefont{Auerbach}},
  \bibinfo{author}{\bibfnamefont{V.~V.} \bibnamefont{Flambaum}},
  \bibnamefont{and} \bibinfo{author}{\bibfnamefont{V.}~\bibnamefont{Spevak}},
  \bibinfo{journal}{Phys. Rev. Lett.} \textbf{\bibinfo{volume}{76}},
  \bibinfo{pages}{4316} (\bibinfo{year}{1996}).

\bibitem[{\citenamefont{Spevak et~al.}(1997)\citenamefont{Spevak, Auerbach, and
  Flambaum}}]{RaSchiffmoment:02}
\bibinfo{author}{\bibfnamefont{V.}~\bibnamefont{Spevak}},
  \bibinfo{author}{\bibfnamefont{N.}~\bibnamefont{Auerbach}}, \bibnamefont{and}
  \bibinfo{author}{\bibfnamefont{V.~V.}~\bibnamefont{Flambaum}},
  \bibinfo{journal}{Phys. Rev. C} \textbf{\bibinfo{volume}{56}},
  \bibinfo{pages}{1357} (\bibinfo{year}{1997}).

\bibitem[{\citenamefont{Sushkov
  et~al.}(1984{\natexlab{a}})\citenamefont{Sushkov, Flambaum, and
  Khriplovich}}]{TlSchiffmoment:01}
\bibinfo{author}{\bibfnamefont{O.~P.} \bibnamefont{Sushkov}},
  \bibinfo{author}{\bibfnamefont{V.~V.} \bibnamefont{Flambaum}},
  \bibnamefont{and} \bibinfo{author}{\bibfnamefont{I.~B.}
  \bibnamefont{Khriplovich}}, \bibinfo{journal}{Zh. Eksp. Teor. Fiz.}
  \textbf{\bibinfo{volume}{87}}, \bibinfo{pages}{1521}
  (\bibinfo{year}{1984}{\natexlab{a}}).

\bibitem[{\citenamefont{Sushkov
  et~al.}(1984{\natexlab{b}})\citenamefont{Sushkov, Flambaum, and
  Khriplovich}}]{TlSchiffmoment:02}
\bibinfo{author}{\bibfnamefont{O.~P.} \bibnamefont{Sushkov}},
  \bibinfo{author}{\bibfnamefont{V.~V.} \bibnamefont{Flambaum}},
  \bibnamefont{and} \bibinfo{author}{\bibfnamefont{I.~B.}
  \bibnamefont{Khriplovich}}, \bibinfo{journal}{Sov. Phys. JETP}
  \textbf{\bibinfo{volume}{60}}, \bibinfo{pages}{873}
  (\bibinfo{year}{1984}{\natexlab{b}}).

\bibitem[{\citenamefont{Flambaum et~al.}(1986)\citenamefont{Flambaum,
  Khriplovich, and Sushkov}}]{TlSchiffmoment:03}
\bibinfo{author}{\bibfnamefont{V.~V.} \bibnamefont{Flambaum}},
  \bibinfo{author}{\bibfnamefont{I.~B.} \bibnamefont{Khriplovich}},
  \bibnamefont{and} \bibinfo{author}{\bibfnamefont{O.~P.}
  \bibnamefont{Sushkov}}, \bibinfo{journal}{Nucl. Phys. A}
  \textbf{\bibinfo{volume}{449}}, \bibinfo{pages}{750} (\bibinfo{year}{1986}).

\bibitem[{\citenamefont{Titov et~al.}(2006)\citenamefont{Titov, Mosyagin,
  Petrov, Isaev, and DeMille}}]{Titov:06amin}
\bibinfo{author}{\bibfnamefont{A.~V.} \bibnamefont{Titov}},
  \bibinfo{author}{\bibfnamefont{N.~S.} \bibnamefont{Mosyagin}},
  \bibinfo{author}{\bibfnamefont{A.~N.} \bibnamefont{Petrov}},
  \bibinfo{author}{\bibfnamefont{T.~A.} \bibnamefont{Isaev}}, \bibnamefont{and}
  \bibinfo{author}{\bibfnamefont{D.~P.} \bibnamefont{DeMille}},
  \bibinfo{journal}{Progr.\ Theor.\ Chem.\ Phys.}
  \textbf{\bibinfo{volume}{B~15}}, \bibinfo{pages}{253} (\bibinfo{year}{2006}).

\bibitem[{\citenamefont{Mosyagin et~al.}(2010)\citenamefont{Mosyagin,
  Zaitsevskii, and Titov}}]{Mosyagin:10a}
\bibinfo{author}{\bibfnamefont{N.~S.} \bibnamefont{Mosyagin}},
  \bibinfo{author}{\bibfnamefont{A.~V.} \bibnamefont{Zaitsevskii}},
  \bibnamefont{and} \bibinfo{author}{\bibfnamefont{A.~V.} \bibnamefont{Titov}},
  \bibinfo{journal}{Review of Atomic and Molecular Physics}
  \textbf{\bibinfo{volume}{1}}, \bibinfo{pages}{63} (\bibinfo{year}{2010}).

\bibitem[{\citenamefont{Titov and Mosyagin}(1999)}]{Titov:99}
\bibinfo{author}{\bibfnamefont{A.~V.} \bibnamefont{Titov}} \bibnamefont{and}
  \bibinfo{author}{\bibfnamefont{N.~S.} \bibnamefont{Mosyagin}},
  \bibinfo{journal}{Int.\ J.\ Quantum Chem.} \textbf{\bibinfo{volume}{71}},
  \bibinfo{pages}{359} (\bibinfo{year}{1999}).

\bibitem[{\citenamefont{Baklanov et~al.}(2010)\citenamefont{Baklanov, Petrov,
  Titov, and Kozlov}}]{Baklanov:10}
\bibinfo{author}{\bibfnamefont{K.~I.} \bibnamefont{Baklanov}},
  \bibinfo{author}{\bibfnamefont{A.~N.} \bibnamefont{Petrov}},
  \bibinfo{author}{\bibfnamefont{A.~V.} \bibnamefont{Titov}}, \bibnamefont{and}
  \bibinfo{author}{\bibfnamefont{M.~G.} \bibnamefont{Kozlov}},
  \bibinfo{journal}{Phys.\ Rev.\ A} \textbf{\bibinfo{volume}{82}},
  \bibinfo{pages}{060501(R)/1} (\bibinfo{year}{2010}).

\bibitem[{\citenamefont{Petrov et~al.}(2007)\citenamefont{Petrov, Mosyagin,
  Isaev, and Titov}}]{Petrov:07a}
\bibinfo{author}{\bibfnamefont{A.~N.} \bibnamefont{Petrov}},
  \bibinfo{author}{\bibfnamefont{N.~S.} \bibnamefont{Mosyagin}},
  \bibinfo{author}{\bibfnamefont{T.~A.} \bibnamefont{Isaev}}, \bibnamefont{and}
  \bibinfo{author}{\bibfnamefont{A.~V.} \bibnamefont{Titov}},
  \bibinfo{journal}{Phys.\ Rev.\ A} \textbf{\bibinfo{volume}{76}},
  \bibinfo{pages}{030501(R)} (\bibinfo{year}{2007}).

\bibitem[{\citenamefont{Skripnikov et~al.}(2009)\citenamefont{Skripnikov,
  Petrov, Titov, and Mosyagin}}]{Skripnikov:09}
\bibinfo{author}{\bibfnamefont{L.~V.} \bibnamefont{Skripnikov}},
  \bibinfo{author}{\bibfnamefont{A.~N.} \bibnamefont{Petrov}},
  \bibinfo{author}{\bibfnamefont{A.~V.} \bibnamefont{Titov}}, \bibnamefont{and}
  \bibinfo{author}{\bibfnamefont{N.~S.} \bibnamefont{Mosyagin}},
  \bibinfo{journal}{Phys.\ Rev.\ A} \textbf{\bibinfo{volume}{80}},
  \bibinfo{pages}{060501(R)} (\bibinfo{year}{2009}).

\bibitem[{\citenamefont{Skripnikov et~al.}(2011)\citenamefont{Skripnikov,
  Titov, Petrov, Mosyagin, and Sushkov}}]{Skripnikov:11a}
\bibinfo{author}{\bibfnamefont{L.~V.}~\bibnamefont{Skripnikov}},
  \bibinfo{author}{\bibfnamefont{A.~V.}~\bibnamefont{Titov}},
  \bibinfo{author}{\bibfnamefont{A.~N.} \bibnamefont{Petrov}},
  \bibinfo{author}{\bibfnamefont{N.~S.} \bibnamefont{Mosyagin}},
  \bibnamefont{and} \bibinfo{author}{\bibfnamefont{O.~P.}
  \bibnamefont{Sushkov}}, \bibinfo{journal}{Phys.\ Rev.\ A}
  \textbf{\bibinfo{volume}{84}}, \bibinfo{pages}{022505}
  (\bibinfo{year}{2011}).

\bibitem[{\citenamefont{Petrov}(2011)}]{Petrov:11}
\bibinfo{author}{\bibfnamefont{A.~N.} \bibnamefont{Petrov}},
  \bibinfo{journal}{Phys.\ Rev.\ A} \textbf{\bibinfo{volume}{83}},
  \bibinfo{pages}{024502} (\bibinfo{year}{2011}).

\bibitem[{\citenamefont{Kaldor et~al.}()\citenamefont{Kaldor, Eliav, and
  Landau}}]{MolRCCSD}
\bibinfo{author}{\bibfnamefont{U.}~\bibnamefont{Kaldor}},
  \bibinfo{author}{\bibfnamefont{E.}~\bibnamefont{Eliav}}, \bibnamefont{and}
  \bibinfo{author}{\bibfnamefont{A.}~\bibnamefont{Landau}},
  \bibinfo{note}{program package for calculation of molecules by the
  relativistic {F}ock-space coupled-cluster method}.

\bibitem[{\citenamefont{Kaldor et~al.}(2004)\citenamefont{Kaldor, Eliav, and
  Landau}}]{Kaldor:04ba}
\bibinfo{author}{\bibfnamefont{U.}~\bibnamefont{Kaldor}},
  \bibinfo{author}{\bibfnamefont{E.}~\bibnamefont{Eliav}}, \bibnamefont{and}
  \bibinfo{author}{\bibfnamefont{A.}~\bibnamefont{Landau}}, in
  \emph{\bibinfo{booktitle}{Recent Advances in Relativistic Molecular Theory}},
  edited by \bibinfo{editor}{\bibfnamefont{K.}~\bibnamefont{Hirao}}
  \bibnamefont{and} \bibinfo{editor}{\bibfnamefont{Y.}~\bibnamefont{Ishikawa}}
  (\bibinfo{organization}{World Scientific}, \bibinfo{address}{Singapore},
  \bibinfo{year}{2004}), p. \bibinfo{pages}{283}.

\bibitem[{\citenamefont{Stanton et~al.}(2011)\citenamefont{Stanton, Gauss,
  Harding, Szalay et~al.}}]{CFOUR}
\bibinfo{author}{\bibfnamefont{J.~F.} \bibnamefont{Stanton}},
  \bibinfo{author}{\bibfnamefont{J.}~\bibnamefont{Gauss}},
  \bibinfo{author}{\bibfnamefont{M.~E.} \bibnamefont{Harding}},
  \bibinfo{author}{\bibfnamefont{P.~G.} \bibnamefont{Szalay}},
  \bibnamefont{et~al.} (\bibinfo{year}{2011}), \bibinfo{note}{{\sc cfour}: a
  program package for performing high-level quantum chemical calculations on
  atoms and molecules, {http://www.cfour.de} .}

\bibitem[{\citenamefont{K\'{a}llay and Surj\'{a}n}(2001)}]{Kallay:1}
\bibinfo{author}{\bibfnamefont{M.}~\bibnamefont{K\'{a}llay}} \bibnamefont{and}
  \bibinfo{author}{\bibfnamefont{P.~R.} \bibnamefont{Surj\'{a}n}},
  \bibinfo{journal}{J.\ Chem.\ Phys.} \textbf{\bibinfo{volume}{115}},
  \bibinfo{pages}{2945} (\bibinfo{year}{2001}).

\bibitem[{\citenamefont{K\'{a}llay et~al.}(2003)\citenamefont{K\'{a}llay,
  Gauss, and Szalay}}]{Kallay:3}
\bibinfo{author}{\bibfnamefont{M.}~\bibnamefont{K\'{a}llay}},
  \bibinfo{author}{\bibfnamefont{J.}~\bibnamefont{Gauss}}, \bibnamefont{and}
  \bibinfo{author}{\bibfnamefont{P.~G.} \bibnamefont{Szalay}},
  \bibinfo{journal}{J.\ Chem.\ Phys.} \textbf{\bibinfo{volume}{119}},
  \bibinfo{pages}{2991} (\bibinfo{year}{2003}).

\bibitem[{\citenamefont{Mosyagin et~al.}(2000)\citenamefont{Mosyagin, Eliav,
  Titov, and Kaldor}}]{Mosyagin:00}
\bibinfo{author}{\bibfnamefont{N.~S.} \bibnamefont{Mosyagin}},
  \bibinfo{author}{\bibfnamefont{E.}~\bibnamefont{Eliav}},
  \bibinfo{author}{\bibfnamefont{A.~V.} \bibnamefont{Titov}}, \bibnamefont{and}
  \bibinfo{author}{\bibfnamefont{U.}~\bibnamefont{Kaldor}},
  \bibinfo{journal}{J.\ Phys.\ B} \textbf{\bibinfo{volume}{33}},
  \bibinfo{pages}{667} (\bibinfo{year}{2000}).

\bibitem[{\citenamefont{Rasmussen}(1934)}]{Rasmuss:01}
\bibinfo{author}{\bibfnamefont{E.}~\bibnamefont{Rasmussen}},
  \bibinfo{journal}{Zeit. Phys.} \textbf{\bibinfo{volume}{87}},
  \bibinfo{pages}{607} (\bibinfo{year}{1934}).

\bibitem[{\citenamefont{Russel}(1934)}]{Russell:01}
\bibinfo{author}{\bibfnamefont{H.~N.} \bibnamefont{Russel}},
  \bibinfo{journal}{Phys. Rev.} \textbf{\bibinfo{volume}{46}},
  \bibinfo{pages}{989} (\bibinfo{year}{1934}).

\bibitem[{\citenamefont{Dzuba et~al.}(2002)\citenamefont{Dzuba, Flambaum,
  Ginges, and Kozlov}}]{Dzuba:02}
\bibinfo{author}{\bibfnamefont{V.~A.} \bibnamefont{Dzuba}},
  \bibinfo{author}{\bibfnamefont{V.~V.} \bibnamefont{Flambaum}},
  \bibinfo{author}{\bibfnamefont{J.~S.~M.} \bibnamefont{Ginges}},
  \bibnamefont{and} \bibinfo{author}{\bibfnamefont{M.~G.}
  \bibnamefont{Kozlov}}, \bibinfo{journal}{Phys.\ Rev.\ A}
  \textbf{\bibinfo{volume}{66}}, \bibinfo{pages}{012111}
  (\bibinfo{year}{2002}).

\bibitem[{\citenamefont{Flambaum and Khriplovich}(1984)}]{Z-scaling:01}
\bibinfo{author}{\bibfnamefont{V.~V.} \bibnamefont{Flambaum}} \bibnamefont{and}
  \bibinfo{author}{\bibfnamefont{I.~B.} \bibnamefont{Khriplovich}},
  \bibinfo{journal}{Sov. Phys. JETP} \textbf{\bibinfo{volume}{60}},
  \bibinfo{pages}{873} (\bibinfo{year}{1984}).

\bibitem[{\citenamefont{Griffith et~al.}(2009)\citenamefont{Griffith, Swallows,
  Loftus, Romalis, Heckel, and Fortson}}]{Griffith:09}
\bibinfo{author}{\bibfnamefont{W.~C.} \bibnamefont{Griffith}},
  \bibinfo{author}{\bibfnamefont{M.~D.} \bibnamefont{Swallows}},
  \bibinfo{author}{\bibfnamefont{T.~H.} \bibnamefont{Loftus}},
  \bibinfo{author}{\bibfnamefont{M.~V.} \bibnamefont{Romalis}},
  \bibinfo{author}{\bibfnamefont{B.~R.} \bibnamefont{Heckel}},
  \bibnamefont{and} \bibinfo{author}{\bibfnamefont{E.~N.}
  \bibnamefont{Fortson}}, \bibinfo{journal}{Phys. Rev. Lett.}
  \textbf{\bibinfo{volume}{102}}, \bibinfo{pages}{101601}
  (\bibinfo{year}{2009}).

\bibitem[{\citenamefont{Cho et~al.}(1991)\citenamefont{Cho, Sangster, and
  Hinds}}]{Cho:91}
\bibinfo{author}{\bibfnamefont{D.}~\bibnamefont{Cho}},
  \bibinfo{author}{\bibfnamefont{K.}~\bibnamefont{Sangster}}, \bibnamefont{and}
  \bibinfo{author}{\bibfnamefont{E.~A.} \bibnamefont{Hinds}},
  \bibinfo{journal}{Phys.\ Rev.\ A} \textbf{\bibinfo{volume}{44}},
  \bibinfo{pages}{2783} (\bibinfo{year}{1991}).

\end{thebibliography}
\end{document}